\documentclass[12pt,preprint]{aastex}
\slugcomment{To appear in the Astrophysical Journal.}
\slugcomment{\bf DRAFT: \today}
\shorttitle{Potential \ion{N}{1} Variations}
\shortauthors{Knauth et al.}

\begin{document}

\title{Potential Variations in the Interstellar \ion{N}{1} 
Abundance\altaffilmark{1}}

\author{David C. Knauth\altaffilmark{2},
B-G Andersson\altaffilmark{2}, Stephan R. McCandliss\altaffilmark{2},
and H. Warren Moos\altaffilmark{2}}

\altaffiltext{1}{Based on observations made with the NASA-CNES-CSA $Far$
$Ultraviolet$ $Spectroscopic$ $Explorer$, which is operated for NASA 
by the Johns Hopkins University under NASA contract NAS 5-32985, and the 
NASA/ESA $Hubble$ $Space$ $Telescope$, obtained from the MAST data archive at 
the Space Telescope Science Institute. STScI is operated by the Association of 
Universities for Research in Astronomy, Inc. under the NASA contract 
NAS 5-26555.}

\altaffiltext{2}{Department of Physics and Astronomy, The Johns Hopkins 
University, Baltimore, MD 21218; dknauth@pha.jhu.edu.}
{}

\vfill

\begin{abstract}
We present {\it Far Ultraviolet Spectroscopic Explorer (FUSE)} and Space
Telescope Imaging Spectrograph observations of the weak interstellar
N~{\small I} $\lambda$1160 doublet toward 17 high-density sightlines 
[$N$(H$_{tot}$) $\geq$ 10$^{21}$ cm$^{-2}$].  When combined with published data,
our results reveal variations in the fractional \ion{N}{1} abundance showing a 
systematic deficiency at large $N$(H$_{tot}$).  At the {\it FUSE} resolution
($\sim$ 20 km s$^{-1}$) the effects of unresolved saturation cannot
be conclusively ruled out, although \ion{O}{1} $\lambda$1356 shows little 
evidence for saturation.  We investigated the possibility that the \ion{N}{1}
variability is due to the formation of N$_2$ in our mostly dense regions.  The 
0-0 band of the c$^{\prime}_{4}$$^1$$\Sigma$$_u^+$ $-$ X$^1$$\Sigma$$_g^+$ 
transition of N$_2$ at 958 \AA\ should be easily detected in our $FUSE$ data; 
for 10 of the denser sightlines N$_2$ is not observed at a sensitivity level of 
a few times 10$^{14}$ cm$^{-2}$.  The observed \ion{N}{1} variations 
are suggestive of an incomplete understanding of nitrogen chemistry.
\end{abstract}


\section{Introduction}
Elemental abundance studies are important for models of Galactic chemical 
evolution.  Studies of the interstellar oxygen abundance support a relatively
constant value of O/H$_{tot}$ = (3.43 $\pm$ 0.15) $\times$ 10$^{-4}$ out to 
$\sim$ 1000 kpc, with any variability less than the 1 $\sigma$ measurement 
uncertainties (Meyer, Jura, \& Cardelli 1998; Cartledge et al. 2001;
Andr\'{e} et al. 2003).  The situation is not as clear for carbon and 
nitrogen.  In this work, we focus specifically on the abundance of interstellar
nitrogen.   

It is generally believed that interstellar nitrogen is a product of the CNO 
cycle and recycled into the ISM through the winds of low and intermediate mass
stars (e.g., red giant branch and/or asymptotic giant branch stars (AGB);
Pilyugin, Thuan, \& Vichez 2003).  In contrast, oxygen is produced in stars 
during the He burning phase and is returned to the ISM via supernovae.  

Early studies (Lugger et al. 1978; Ferlet 1981; York et al. 1983) of 
interstellar N~{\small I} using the {\it Copernicus} satellite investigated 
moderately dense lines of sight and found that $N$(\ion{N}{1}) increases 
linearly with $N$(H$_{tot}$) [$N$(H$_{tot}$) = 2$N$(H$_2$) + $N$(\ion{H}{1})].  
Using high-quality Goddard High-Resolution Spectrograph (GHRS) data for seven 
sight lines with $N$(H$_{tot}$) $\leq$ 10$^{21}$ cm$^{-2}$, Meyer, Cardelli, \& 
Sofia (1997) suggested that the interstellar nitrogen abundance is constant, 
N/H$_{tot}$ = (7.5 $\pm$ 0.4) $\times$ 10$^{-5}$.  However, Jenkins
et al. (1999) and Sonneborn et al. (2000), using data from the Interstellar 
Medium Absorption Profile Spectrograph (IMAPS), show a factor-of-2 variation in
the N~{\small I} abundance between $\delta$~Ori [\ion{N}{1}/H$_{tot}$ = (3.97 
$\pm$ 0.30) $\times$ 10$^{-5}$] and $\gamma^2$ Vel [\ion{N}{1}/H$_{tot}$ = 
(7.99 $\pm$ 0.47) $\times$ 10$^{-5}$].  

In the near interstellar medium (ISM), $d$ $\lesssim$ 100pc, \ion{N}{1} is more
affected by ionization than \ion{O}{1} because \ion{O}{1} is coupled more 
strongly to \ion{H}{1} by charge exchange reactions (Sofia \& Jenkins 1998;
Jenkins et al. 2000; Lehner et al. 2003).  Therefore, these near ISM sightlines
(Lehner et al. 2003) will not be discussed here.  The effects of ionization at 
larger column densities are thought to be insignificant.  In addition, nitrogen
is not incorporated into refractory interstellar dust grains (Sofia, Cardelli, 
\& Savage 1994).  Nitrogen bearing interstellar ices (Gibb, Whittet, \& Chiar 
2001; Chiar et al. 2002) are not expected along our moderately reddened ($A_v$ 
$\lesssim$ 2.0) sight lines since interstellar ices do not form until $A_v$ 
$\geq$ 3 (Whittet et al. 2001). However, nitrogen chemistry and N$_2$ formation
should become important.

In order to probe the extent of potential variations in the interstellar 
nitrogen abundance, we have undertaken a survey of the weak interstellar 
\ion{N}{1} doublet $\lambda\lambda$1159.817, 1160.937.  Use of this doublet
removes uncertainty in oscillator strengths ($f$-values) for different 
transitions and allows for a direct comparison to the work of Meyer et al. 
(1997).  Our survey utilized archival data from the {\it Far Ultraviolet 
Spectroscopic Explorer (FUSE; {\rm Moos et al. 2000})} and the Space Telescope
Imaging Spectrograph (STIS) on board the {\it Hubble Space Telescope (HST)} 
toward 17 high-density [$N$(H) $\geq$ 10$^{21}$ cm $^{-2}$] sightlines.

\section{Observations and Data Reduction}
\subsection{$FUSE$ Data}
The \ion{N}{1} data for 16 of the stars studied here, shown in Table~\ref{N_H},
were obtained throughout the $FUSE$ Prime mission phase (November 1999 $-$ March
2003).  All of the data were acquired with the star in the large aperture 
(30~$\!\!^{\prime\prime}$ x 30~$\!\!^{\prime\prime}$; LWRS) with the exception
of HD~219188 for which the medium aperture (4~$\!\!^{\prime\prime}$ x 
20~$\!\!^{\prime\prime}$; MDRS) was utilized.  The data cover the
wavelength range 905 $-$ 1185 \AA\ with a spectral resolution equivalent to 
$\Delta$$v$ $\sim$ 20 km s$^{-1}$.  The weak interstellar \ion{N}{1} doublet
at 1160 \AA\ appears in both the LiF1 and LiF2 channels which provides a 
consistency check to rule out the possibility of detector artifacts.

The time-tagged and histogram data were reduced and calibrated with 
CalFUSE\footnote{The CalFUSE Pipeline Reference Guide is available at 
http://fuse.pha.jhu.edu/analysis/pipeline\_reference.html.} (version 2.2.2;
Dixon \& Sahnow 2003).  CalFUSE provides the appropriate Doppler corrections 
to remove the effects of the spacecraft motion and places the data on the 
heliocentric velocity scale ($V_{helio}$). The wavelength solution provides 
good relative calibration across the LiF channels.  In order to minimize the
uncertainties in the relative wavelength calibration between exposures, the 
data were co-added with a cross correlation technique.  We used interstellar 
H$_2$ lines in several of the $FUSE$ channels to fix the velocity scale 
($V_{rel}$).  The final summed spectra (near 1160 \AA) have signal-to-noise 
(S/N) ratios between $\sim$~20 and $\sim$~300 per resolution element for all
data sets.   S/N ratios greater than 30 were obtained utilizing focal plane 
splits or other similar procedures. 

\subsection{STIS Data}
STIS observed HD 147888 (\ion{N}{1} $\lambda$1161) for 1656 s on 2000 Aug 17, 
HD~24534 (\ion{O}{1} $\lambda$1356) for 2945 s on 2001 Mar 25, and HD~219188 
(\ion{O}{1} $\lambda$1356) for 1200 s on 2001 Sept 27.  The \ion{N}{1} 
$\lambda$1161 data for HD~147888 used the E140H grating centered
at 1271 \AA\ and the 0.2$^{\prime\prime}$ x 0.09$^{\prime\prime}$ aperture 
resulting in a resolving power ($R$) of about 100,000 ($\Delta$$v$ $\sim$ 2.8 
km s$^{-1}$).  The \ion{O}{1} $\lambda$1356 data utilized
the same grating but with the 0.1$^{\prime\prime}$ x 0.03$^{\prime\prime}$
aperture centered on $\lambda$1416 for HD~24534 and on $\lambda$1271 for 
HD~219188 resulting in $R$ $\sim$ 200,000 ($\Delta$$v$ $\sim$ 1.5 km s$^{-1}$).

The E140H setup centered at $\lambda$1271 used to measure \ion{N}{1} toward 
HD~147888, is not ideal since the stronger \ion{N}{1} line is not obtained.  In 
addition, the MgF$_2$ detector windows severely attenuates the flux below 1200 
\AA.  The 1161 \AA\ line can only be detected in relatively dense 
sightlines or with long exposure times.  With the exception of HD~147888, 
examination of the archival data for other sightlines did not reveal evidence
for the \ion{N}{1} 1161 \AA\ line due to the relatively low S/N ($\leq$ 15 per
resolution element).   

The data were reduced and extracted with the CALSTIS pipeline (v2.13b).  The 
subtraction of background and scattered light from the echelle data employs 
the algorithm of Lindler \& Bowers (2000).  CALSTIS provides the 
appropriate Doppler corrections to place the spectra on the $V_{helio}$ scale. 
The spectrum of HD~147888 at \ion{N}{1} $\lambda$1161 has a S/N ratio of 
$\sim$ 12 per resolution element.  For \ion{O}{1} $\lambda$1356 toward 
HD~24534 and HD~219188, we obtained a S/N of $\sim$ 70 and 48 per resolution 
element, respectively.

\section{Results and Discussion}
\begin{figure}\figurenum{1}\epsscale{0.55}
\plotone{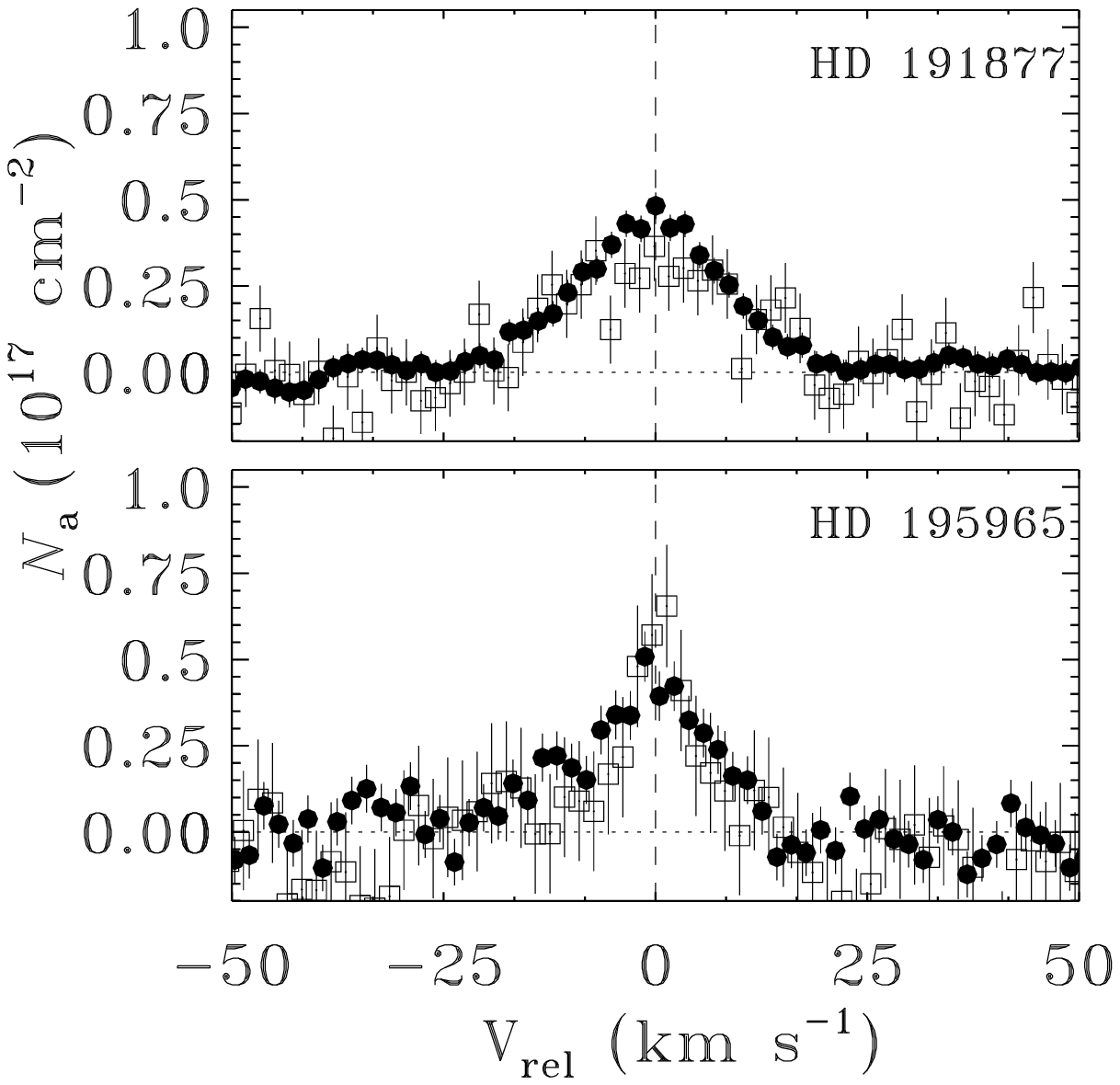}
\vspace{0.75in} 
\caption{\label{Na} $Top$: N I apparent column density profiles for 
HD~191877 and HD~195965.  The stronger line is denoted
by the solid circles, while the weaker line is represented by the open squares.
The vertical lines represent 1 $\sigma$ uncertainties that 
were determined following the procedures outlined in Sembach \& Savage (1992). 
The signal-to-noise ratios are 320 per resolution element for HD~191877 
and 270 per resolution element for HD~195965.  There is no evidence for 
unresolved saturated structure.}
\end{figure}

Figure~\ref{Na} shows the apparent column density (ACD) profiles, $N_a(v)$, for 
both members of the weak interstellar \ion{N}{1} doublet toward 2 stars of our 
stellar sample.  See Savage \& Sembach (1991) for a detailed 
discussion of the apparent optical depth method.  The excellent overall 
agreement between the $N_a(v)$ profiles for both members of the \ion{N}{1} 
doublet implies that the apparent column densities are the true column 
densities.  Table~\ref{N_H} presents our \ion{N}{1} column densities, the 
average of both members of the doublet except HD~147888 (assumed to 
be unsaturated).  Also presented are the measured equivalent widths 
($W_{\lambda}$s) of the \ion{N}{1} doublet for each star.  Finally, 
Table~\ref{N_H} shows the available $Copernicus$, $International$ $Ultraviolet$
$Explorer$, GHRS, IMAPS, and STIS measurements of $N$(\ion{N}{1}), 
$N$(H$_{tot}$), and $N$(\ion{O}{1}).  

\begin{figure}\figurenum{2}\epsscale{0.65}
\plotone{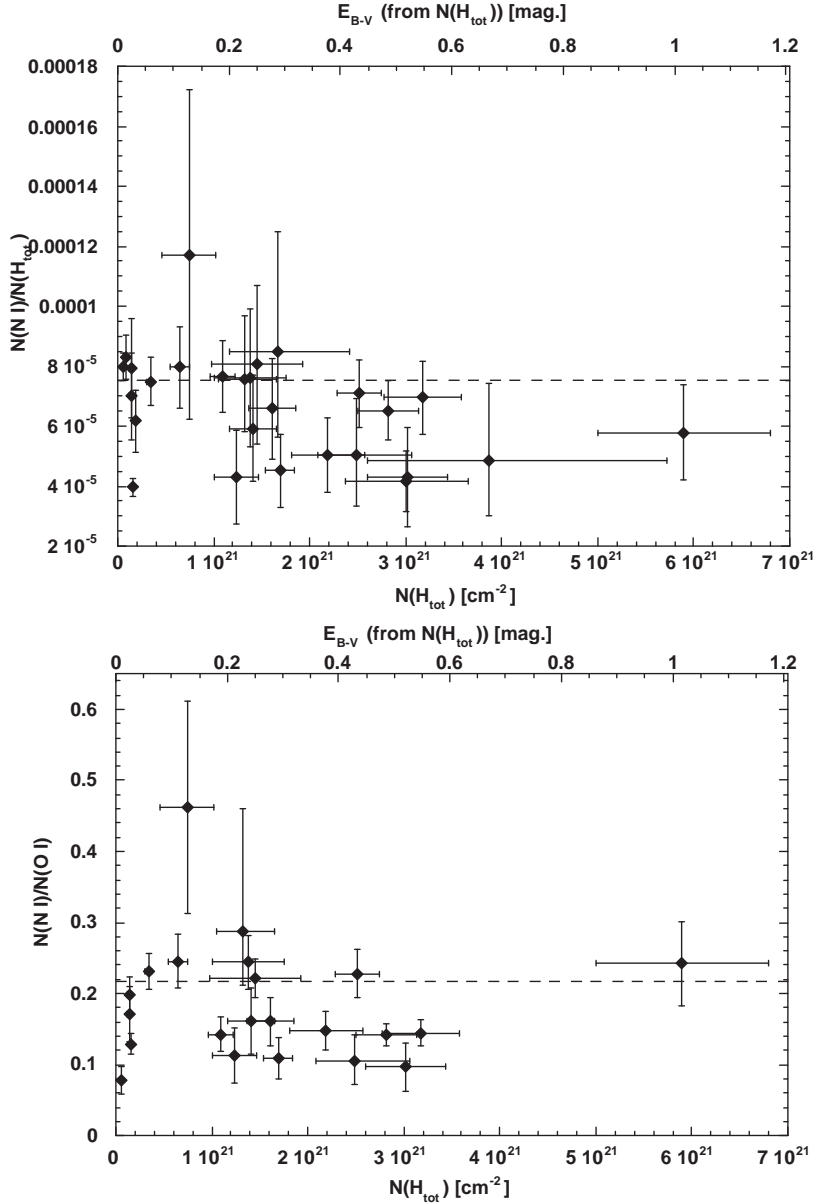}
\vspace{0.15in} 
\caption{\label{NvsH} $Top$: $N$(N I)/$N$(H$_{tot}$) as a function
of $N$(H$_{tot}$).  The dashed line represents the
average value N/H$_{tot}$ from Meyer et al. (1997).
For $N$(H$_{tot}$) $\geq$ 10$^{21}$ cm$^{-2}$ there is a large dispersion from
the average, unlike that found for interstellar O I (e.g., Meyer et al.
1998).  The observed variability either represents a real cosmic variance or 
evidence for enhanced chemical processing of N I.  $Bottom$: 
$N$(N I)/$N$(O I) as a function of $N$(H$_{tot}$).  The 
dashed line is Meyer's average values for N/O [i.e., 
(N/H$_{tot}$)/(O/H$_{tot}$); Meyer et al. 1997; Meyer et al. 1998]. The large
scatter verifies that N I varies with $N$(H$_{tot}$) and is not an 
artifact of the larger uncertainties associated with $N$(H$_{tot}$).}
\end{figure}

Figure~\ref{NvsH} ($Top$) shows $N$(\ion{N}{1})/$N$(H$_{tot}$) versus
$N$(H$_{tot}$) which reveals a departure from the average
at high column densities.  With the exception of $\delta$~Ori, the \ion{N}{1} 
abundance appears to be constant for $N$(H$_{tot}$) $\leq$ 10$^{21}$ cm$^{-2}$,
but an anticorrelation seems to be present for greater column densities.  A 
Student t test yields a 7.6\% probability that the data sets for $N$(H$_{tot}$)
$\leq$ 10$^{21}$ cm$^{-2}$ and $N$(H$_{tot}$) $\geq$ 10$^{21}$ cm$^{-2}$ 
(including the outlier $\delta$~Ori) arise from the same parent population.  If
$\delta$~Ori is excluded, the probability is reduced to 1.3\%.  Therefore, we
conclude that the observed variability is real.  Figure~\ref{NvsH} ($Bottom$)
depicts $N$(\ion{N}{1})/$N$(\ion{O}{1}) as a function of $N$(H$_{tot}$).  The
data point for HD~147888, the densest sightline presented here, is high because 
\ion{O}{1} is slightly depleted (Cartledge et al. 2001).  The large scatter in 
N/O supports that the variation in \ion{N}{1} is not an artifact introduced by 
the larger uncertainties for $N$(H$_{tot}$).  

In order to test whether the observed \ion{N}{1} variability is due to 
observational effects (i.e., saturation) we calculated curves-of-growth for
four stars which exhibited deviant behavior (e.g., HD~179406) and six stars
which did not.  The curve of growth results agree with those obtained from the
ACD method.    Additionally, our results compare favorably to previous 
\ion{N}{1} measurements (Hoopes et al. 2003; Sonnentrucker 2003, private 
communication).  We believe saturation effects to be minimal since 
$\tau_{{\rm N~I}}$/$\tau_{{\rm O~I}}$ = 1.4 and \ion{O}{1} shows little
evidence for saturation (e.g., Cartledge et al. 2001). 

A second source of systematic error could arise if the instrinsic absorption
profile consisted of a narrow, but unsaturated, feature plus broad shallow 
wings.  In such a scenario, the weak \ion{N}{1} doublet associated with the 
broad component could be too weak to be detected at low S/N.  If enough 
\ion{H}{1} column is located in the broad component, this could skew 
the \ion{N}{1}/H$_{tot}$ ratio to systematically low values.  We can 
investigate this possibility in two ways: 1.) through a comparison of the 
measured Doppler broadening parameter, $b$-value, for the \ion{N}{1} and 
hydrogen lines and 2.) comparing \ion{N}{1}/H$_{tot}$ vs. distance to test 
whether systematically lower ratios are caused by additional weak kinematic 
components for the longer lines-of-sight (Spitzer 1985).  

Unfortunately, both the Ly-$\alpha$ line of H~I and the $J$ = 0, 1 lines of 
H$_2$ typically lie on the square-root part of the curve of growth for these 
sightlines and are insensitive to the $b$-value. However, 
$b$-values can nominally be measured for the $J$ $\geq$ 2 lines of H$_2$. 
The published $b$-values for HD~73882, HD~110432, HD~185418, and HD~192639
(Rachford et al. 2001; Sonnentrucker et al. 2002; Sonnentrucker 2003, 
private communication), all show comparable or 
smaller $b$-values for H$_2$ than for N~I indicating that all the \ion{N}{1}
is detected.  Additionally, no statistically significant slope of 
\ion{N}{1}/H$_{tot}$ vs. stellar distance (125pc $\leq$ $d$ $\leq$ 2kpc) is 
found for our sample. These points argue against intrinsic line shape 
contributing to the \ion{N}{1}/H$_{tot}$ deficiencies reported herein.

\subsection{N$_2$}
Based on models of steady-state, gas-phase interstellar chemistry, N$_2$ 
is expected to be the most abundant nitrogen-bearing molecule in dense
clouds.  Viala (1986) predicts that at the column densities of the sight lines
studied here [$N$(H$_{tot}$) of a few times 10$^{21}$ cm$^{-2}$], the N$_2$ 
column densities should be on the order of 15\% of the \ion{N}{1} abundance or
$N$(N$_2$) $\approx$ 10$^{16}$ cm$^{-2}$.   Inclusion of time dependence 
or depletion onto grain mantles (Bergin, Langer, \& Goldsmith 1995) could 
result in a slightly smaller gas-phase N$_2$ abundance.  However, N$_2$
is still predicted to be the most abundant nitrogen bearing molecule.
  
The strongest band of N$_2$, covered by $FUSE$, is the 0-0 band of the 
c$^{\prime}_{4}$$^1$$\Sigma$$_u^+$ $-$ X$^1$$\Sigma$$_g^+$ transition of N$_2$
at 958 \AA.  Other N$_2$ bands reside in the far-ultraviolet, but are blended 
with more abundant species (e.g., H$_2$).  Utilizing the laboratory wavelengths
and $f$-values for the 0-0 band (Stark et al. 2000), we created several 
synthetic N$_2$ spectra, assuming level populations for excitation temperatures 
between 10 $-$ 1000 K (c.f., McCandliss 2003).  Our synthetic spectrum is
saturated for $N$(N$_2$) = 10$^{16}$ cm$^{-2}$.  Therefore, we searched our
$FUSE$ data for the presence of N$_2$.  

\begin{figure}\figurenum{3}\epsscale{0.65}
\plotone{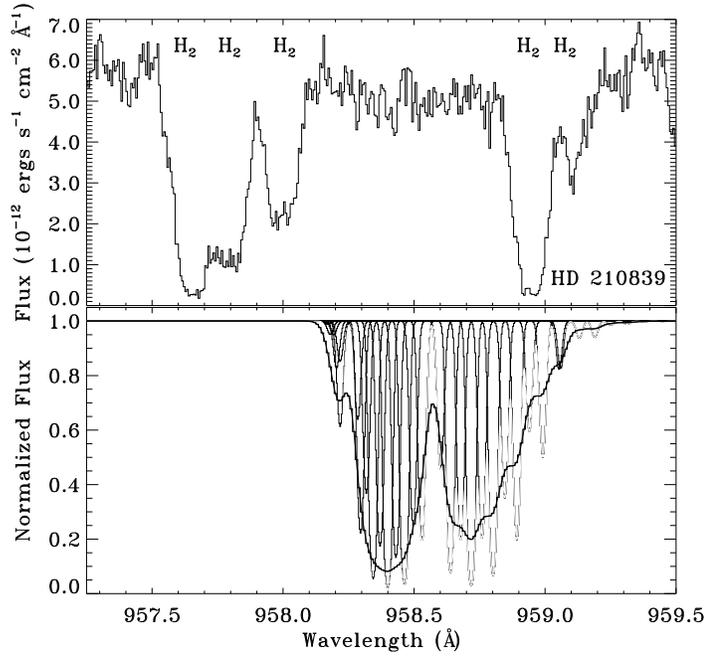}
\vspace{0.3in} 
\caption{\label{N2} $Top$ $-$ The 958 \AA\ portion of the $FUSE$ data toward 
HD~210839 (S/N $\sim$ 30 per resolution element).  $Bottom$ $-$ A synthetic 
spectrum of the 0-0 band of the c$^{\prime}_{4}$ $^1$$\Sigma$$_u^+$ $-$ 
X$^1$$\Sigma$$_g^+$ transition of N$_2$ at 958 \AA, which is the strongest 
N$_2$ band in the $FUSE$ wavelength region.  The column density of the 
synthetic spectrum is $N$(N$_2$) = 10$^{15}$ cm$^{-2}$, 10\% of that expected
from steady-state gas-phase chemistry (Viala 1986).  The synthetic spectrum 
was produced at $T$ = 100 K, $b$ = 5 km s$^{-1}$ (narrow histograms) and 
convolved to the $FUSE$ instrumental resolution $\sim$ 17,000 (solid line).  
N$_2$ should be easily detected in the $FUSE$ data but it is not.}
\end{figure}

Figure~\ref{N2} ($Top$) exhibits the 958 \AA\ portion of the spectrum toward 
HD~210839 [$E$($B$ $-$ $V$) = 0.62 mag.] and the bottom panel shows our 
synthetic N$_2$ spectrum for $N$(N$_2$) = 10$^{15}$ cm$^{-2}$, 10\% of that 
predicted.  The lower column density incorporates the effects of four velocity
components detected in \ion{O}{1} $\lambda$1356 (Andr\'{e} et al. 2003) along 
the line of sight with similar amounts of material.  There is no evidence for 
N$_2$ in the spectrum of HD~210839 to a level of a few times 10$^{14}$
cm$^{-2}$.  Examination of the ten densest sightlines that have $FUSE$ data 
also show a similar dearth of N$_2$.  Hence, gas-phase N$_2$ cannot explain 
the observed \ion{N}{1} variability.  Quantitative limits on N$_2$ will be 
presented in a subsequent paper.  
 
\section{Summary}
Our data show that the interstellar \ion{N}{1} abundances relative to \ion{H}{1}
and \ion{O}{1} appear to vary.  Very high-resolution studies of these and other
sight lines will help to verify and expand the findings reported here.  Although
N$_2$ is expected to be the most abundant nitrogen-bearing molecule in 
moderately dense [log $N$(H$_2$) $\geq$ 10$^{21}$ cm$^{-2}$] clouds (e.g., 
Viala 1986), our data do not support the predicted abundances.  
The presence of an anticorrelation above $N$(H$_{tot}$) = 10$^{21}$ cm$^{-2}$
suggests that the mostly likely explanation of the interstellar \ion{N}{1}
variability is the need for a better understanding of nitrogen chemistry.
However, differences in the mixing processes of products from AGB stellar
outflows and SN may also contribute.   

\acknowledgments We thank the {\it FUSE} Science Operations team for their
dedication and efforts in acquiring these observations.  This research made 
use of the Simbad database, operated at CDS, Strasbourg, France.

\newpage

\setcounter{table}{0}
\begin{deluxetable}{ccccccccc}
\tabletypesize{\scriptsize}
\setlength{\tabcolsep}{0.05in} 
\tablecolumns{8}
\tablewidth{0pt}
\tablecaption{ Interstellar N I, O I, 
	and N I/O I Ratios$^1$ \label{N_H} } 
\tablehead{ 
\colhead{Star} & \colhead{$W_{\lambda1160}$\tablenotemark{2} } & 
\colhead{$W_{\lambda1161}$\tablenotemark{2} } & 
\colhead{$N$(N I)} &  \colhead{$N$(N I)/$N$(H$_{tot}$)} &
\colhead{$N$(O I)} & \colhead{N I/O I} & \colhead{Ref} \\
\colhead{} & \colhead{[m\AA]} &\colhead{[m\AA]} &\colhead{[cm$^{-2}$]} &
\colhead{} & \colhead{[cm$^{-2}$]} & \colhead{} & \colhead{} & \colhead{} }
\startdata
$\gamma$$^2$ Vel & $\ldots$ & $\ldots$ & (4.10 $\pm$ 0.21) $\times$ 10$^{15}$ 
 & (7.99 $\pm$ 0.47) $\times$ 10$^{-5}$ & (5.23 $\pm$ 1.27) $\times$ 10$^{16}$ 
 & 0.08 $\pm$ 0.02 & 3, 4 \\
$\zeta$ Pup & $\ldots$& $\ldots$& (7.62 $\pm$ 0.51) $\times$ 10$^{15}$
 & (8.30 $\pm$ 0.74) $\times$ 10$^{-5}$ & $\leq$ 17.70 & $\geq$ 0.02 &
 3, 5 \\
$\iota$~Ori & 1.00 $\pm$ 0.15 & $\leq$ 0.30 & (9.87 $\pm$ 1.48) $\times$ 10$^{15}$ 
 & (7.00 $\pm$ 1.45) $\times$ 10$^{-5}$ & (5.8 $\pm$ 1.0) $\times$ 10$^{16}$ 
 & 0.17 $\pm$ 0.04 & 6, 7, 8, 9 \\
$\gamma$~Cas & 1.15 $\pm$ 0.12 & 0.40 $\pm$ 0.12 & (1.15 $\pm$ 0.11) $\times$ 10$^{16}$ 
 & (7.93 $\pm$ 1.66) $\times$ 10$^{-5}$ & (5.8 $\pm$ 0.5) $\times$ 10$^{16}$ 
 & 0.20 $\pm$ 0.03 & 6, 7, 8, 9 \\
$\delta$~Ori & $\ldots$& $\ldots$& (6.19 $\pm$ 0.31) $\times$ 10$^{15}$ & 
 (3.97 $\pm$ 0.30) $\times$ 10$^{-5}$ & (4.8 $\pm$ 0.5) $\times$ 10$^{16}$ 
 & 0.13 $\pm$ 0.02 & 4, 7, 10\\
$\kappa$~Sco & 1.10 $\pm$ 0.15 & 0.40 $\pm$ 0.15 & (1.11 $\pm$ 0.14) $\times$ 10$^{16}$ 
 & (6.17 $\pm$ 1.04) $\times$ 10$^{-5}$ & $\ldots$ & $\ldots$ & 6, 8, 9 \\ 
$\kappa$~Ori & 2.60 $\pm$ 0.15 & 0.65 $\pm$ 0.15 & (2.54 $\pm$ 0.14) $\times$ 10$^{16}$
 & (7.49 $\pm$ 0.80) $\times$ 10$^{-5}$ & (1.1 $\pm$ 0.1) $\times$ 10$^{17}$ 
 & 0.23 $\pm$ 0.03 & 6, 7, 8, 9 \\
$\lambda$~Ori & 4.95 $\pm$ 0.20 & 1.45 $\pm$ 0.20 & (5.15 $\pm$ 0.30) $\times$ 10$^{16}$
 & (7.97 $\pm$ 1.37) $\times$ 10$^{-5}$ & (2.1 $\pm$ 0.3) $\times$ 10$^{17}$ 
 & 0.25 $\pm$ 0.04 & 6, 7, 8, 9  \\
HD~219188 & 8.4 $\pm$ 1.3 & 2.5 $\pm$ 0.8 & (8.72 $\pm$ 2.43) $\times$ 10$^{16}$ & 
 (1.17 $\pm$ 0.55) $\times$ 10$^{-4}$ & (1.89 $\pm$ 0.31) $\times$ 10$^{17}$ 
 &  0.46 $\pm$ 0.15 & 8, 9, 11 \\
HD~195965 & 8.1 $\pm$ 0.4 & 2.4 $\pm$ 0.4 & (8.36 $\pm$ 0.84) $\times$ 10$^{16}$
 & (7.67 $\pm$ 1.19) $\times$ 10$^{-5}$ & (5.89~$^{+~0.87}_{-~0.76}$)
 $\times$ 10$^{17}$ & 0.14~$^{+~0.03}_{-~0.02}$ & 11, 12 \\
HD~88115  & 5.7 $\pm$ 1.1 & 1.3 $\pm$ 0.6 & (5.33 $\pm$ 1.68) $\times$ 10$^{16}$ 
 & (4.30 $\pm$ 1.57) $\times$ 10$^{-5}$ & (4.7 $\pm$ 0.6) $\times$ 10$^{17}$ 
 &  0.11 $\pm$ 0.04 & 11, 13  \\
HD~191877 & 9.3 $\pm$ 0.5 & 3.0 $\pm$ 0.5 & (1.00 $\pm$ 0.11) $\times$ 10$^{17}$ 
 & (7.58~$^{+~2.11}_{-~0.18}$) $\times$ 10$^{-5}$ & (3.47~$^{+~2.03}_{-~0.84}$)
 $\times$ 10$^{17}$ & 0.29~$^{+~0.17}_{-~0.08}$ & 11, 12 \\
$\zeta$~Oph & 7.56 $\pm$ 0.74 & 2.68 $\pm$ 0.99 & (1.05 $\pm$ 0.13) $\times$ 10$^{17}$ 
 & (7.61 $\pm$ 2.30) $\times$ 10$^{-5}$ & (4.3 $\pm$ 0.4) $\times$ 10$^{17}$
 & 0.24 $\pm$ 0.04 & 6, 7, 8, 9 \\
HD~75309  & 8.4 $\pm$ 1.3 & 2.2 $\pm$ 0.8 & (8.37 $\pm$ 2.07) $\times$ 10$^{16}$ 
 & (5.94 $\pm$ 1.78) $\times$ 10$^{-5}$ & (5.2 $\pm$ 0.8) $\times$ 10$^{17}$ 
 & 0.16 $\pm$ 0.05  & 11, 13, 14 \\
$\delta$~Sco & 11.20 $\pm$ 0.30 & 3.30 $\pm$ 0.30 & (1.17 $\pm$ 0.06) $\times$ 10$^{17}$ 
 & (8.07 $\pm$ 2.65) $\times$ 10$^{-5}$ & (5.3 $\pm$ 0.6) $\times$ 10$^{17}$ 
 & 0.22 $\pm$ 0.03 & 5, 6 \\
HD~218915  & 11.9 $\pm$ 1.2 & 2.5 $\pm$ 1.0 & (1.06 $\pm$ 0.22) $\times$ 10$^{17}$ 
 & (6.58 $\pm$ 1.68) $\times$ 10$^{-5}$ & (6.6 $\pm$ 0.4) $\times$ 10$^{17}$ 
 & 0.16 $\pm$ 0.04 & 11, 13 \\
HD~110432 & 10.5 $\pm$ 0.7 & 4.8 $\pm$ 0.9 & (1.42 $\pm$ 0.21) $\times$ 10$^{17}$ 
 & (8.50~$^{+~3.97}_{-~2.89}$)$\times$ 10$^{-5}$ & $\ldots$ & $\ldots$ 
 &  11, 15 \\	
HD~94493  & 7.7 $\pm$ 1.2 & 2.1 $\pm$ 0.7 & (7.64 $\pm$ 1.93) $\times$ 10$^{16}$ 
 & (4.52 $\pm$ 1.21) $\times$ 10$^{-5}$ & (7.0 $\pm$ 0.6) $\times$ 10$^{17}$ 
 & 0.11 $\pm$ 0.03 & 11, 13 \\
HD~24534 & 10.2 $\pm$ 0.9 & 3.1 $\pm$ 0.8 & (1.10 $\pm$ 0.20) $\times$ 10$^{17}$ 
 & (5.02 $\pm$ 1.25) $\times$ 10$^{-5}$ & (7.43 $\pm$ 0.28) $\times$ 
 10$^{17}$ & 0.15 $\pm$ 0.03 & 11, 15 \\
HD~185418 & 16.0 $\pm$ 2.0 & 2.7 $\pm$ 1.5 & (1.35 $\pm$ 0.37) $\times$ 10$^{17}$ 
 & (5.42~$^{+~1.94}_{-~1.73}$) $\times$ 10$^{-5}$ & (1.20~$^{+~0.25}_{-~0.12}$) 
 $\times$ 10$^{18}$ & (0.11~$^{+~0.03}_{-~0.03}$) & 11, 13, 14, 16 \\	
HD~99857  & 15.5 $\pm$ 1.3 & 5.2 $\pm$ 0.9 & (1.78 $\pm$ 0.24) $\times$ 10$^{17}$ 
 & (7.09 $\pm$ 1.14) $\times$ 10$^{-5}$ & (7.8 $\pm$ 0.5) $\times$ 10$^{17}$ & 0.23 $\pm$ 0.03 & 11, 13 \\
HD~210839 & 16.5 $\pm$ 0.7 & 5.4 $\pm$ 0.8 & (1.84 $\pm$ 0.19) $\times$ 10$^{17}$
 & (6.53 $\pm$ 1.00) $\times$ 10$^{-5}$ & (1.30 $\pm$ 0.06) $\times$ 10$^{18}$ & 0.14 $\pm$ 0.02 & 11, 13 \\
HD~179406 & 10.5 $\pm$ 1.4 & 3.8 $\pm$ 1.3 & (1.25 $\pm$ 0.31) $\times$ 10$^{17}$ 
 & (4.15 $\pm$ 1.01) $\times$ 10$^{-5}$ & $\ldots$ & $\ldots$ & 8, 9, 11 \\
HD~192639 & 12.5 $\pm$ 2.7 & 3.3 $\pm$ 1.6 & (1.30 $\pm$ 0.46) $\times$ 10$^{17}$ 
 & (4.31 $\pm$ 1.64) $\times$ 10$^{-5}$ & (1.35 $\pm$ 0.06) $\times$ 10$^{18}$
 & 0.10 $\pm$ 0.03 & 11, 13, 17  \\
HD~124314 & 19.5 $\pm$ 1.2 & 6.5 $\pm$ 1.1 & (2.21 $\pm$ 0.27) $\times$ 10$^{17}$  
 & (6.95 $\pm$ 1.21) $\times$ 10$^{-5}$ & (1.53 $\pm$ 0.08) $\times$ 10$^{18}$ 
 & 0.14 $\pm$ 0.02 & 11, 13 \\
HD~73882 & 16.0 $\pm$ 1.6 & 5.6 $\pm$ 1.6 & (1.88 $\pm$ 0.38) $\times$ 10$^{17}$ &
 (4.87~$^{+~2.54}_{-~1.86}$) $\times$ 10$^{-5}$ & $\ldots$ & $\ldots$ 
 & 11, 15 \\
HD~147888 & $\ldots$ & 7.3 $\pm$ 1.5 & (3.41 $\pm$ 0.78) $\times$ 10$^{17}$ 
 & (5.78 $\pm$ 1.59) $\times$ 10$^{-5}$ & (1.41 $\pm$ 0.11) $\times$ 10$^{18}$
 & 0.24 $\pm$ 0.06 & 11, 14 \\
\enddata
\begin{center}
\footnotesize{References$-$ (1) In order of increasing $N$(H$_{tot}$) 
with 1 $\sigma$ measurement uncertainties; (2) Expected $W_{\lambda1160} \over 
W_{\lambda1161}$ = 3.53 (3) Sonneborn et al. 2000; 
(4) Fitzpatrick \& Spitzer 1994; (5) Keenan, Hibbert, \& Dufton 1985;
(6) Meyer et al. 1997; (7) Meyer et al. (1998) with updated O I $\lambda$1356
$f$-value (Welty et al. 1999); (8) H$_2$ from Savage et al. 1977;
(9) H I weighted mean of Bohlin, Savage, \& Drake (1978) and Diplas \& Savage
(1994); (10) Jenkins et al. 1999; (11) This work; (12) Hoopes et al. 2003;
(13) Andr\'{e} et al. 2003; (14) Cartledge et al. 2001;
(15) Rachford et al. 2002; (16) Sonnentrucker 2003, private communication;
(17) Sonnentrucker et al. 2002.} 
\end{center} 
\end{deluxetable} 
\clearpage
\end{document}